\documentclass[preprint,12pt,number]{elsarticle}
\usepackage{natbib}
\usepackage{amsmath}   
\usepackage{graphicx}  
\usepackage{amssymb}
\usepackage{lineno}

\begin{document}

\def\be{\begin{equation}}
\def\ee{\end{equation}}
\def\bea{\begin{eqnarray}}
\def\eea{\end{eqnarray}}
\def\bml{\begin{mathletters}}
\def\eml{\end{mathletters}}
\def\l{\label}
\def\b{\bullet}
\def\eqn#1{(~\ref{eq:#1}~)}
\def\no{\nonumber}
\def\av#1{{\langle  #1 \rangle}}
\begin{frontmatter}



\title{Time to fixation in the presence of recombination}


\author{Kavita Jain}

\address{Theoretical Sciences Unit and  Evolutionary and 
Organismal Biology Unit, \\
Jawaharlal Nehru Centre for Advanced Scientific Research, \\Jakkur P.O.,
Bangalore
560064, India} 

\begin{abstract}
We study the evolutionary dynamics of a haploid population
of infinite size recombining with a probability $r$ in a two locus
model. Starting from a low fitness locus, the population is evolved
under mutation, selection and recombination until a finite 
fraction of the population reaches the fittest locus. An analytical method is
developed to calculate the fixation time $T$ to the fittest locus for
various choices of epistasis. We 
find that (1) for negative epistasis, $T$ decreases slowly for small
$r$ but decays fast at larger $r$ (2) for positive epistasis, $T$ increases linearly for small $r$
and mildly for large $r$ (3) for compensatory mutation, $T$ diverges
as a power law with logarithmic corrections as the recombination
fraction approaches a critical value. Our 
calculations are seen to be in good agreement with the exact numerical
results. 
\end{abstract}

\begin{keyword}
fixation time \sep recombination \sep epistasis
\end{keyword}

\end{frontmatter}

\linenumbers

\section{Introduction}
\label{Intro}

Sexual reproduction is ubiquitous in nature - most eukaryotes reproduce
sexually and genetic mixing is common in some bacterias
\cite{Otto:2002, Rice:2002}. However, asexual reproduction 
is not entirely absent. Microbes such as virus and bacteria
reproduce asexually most 
of the time, ancient asexuals \cite{Judson:1996} which have remained
exclusively asexual 
for millions of years persist and human mitochondrial DNA has not
recombined for a few million years \cite{Loewe:2006}. 
It is then natural to ask: under what conditions is one
or the other mode of reproduction preferred?

A detailed study of theoretical models has been helpful in identifying
some relevant parameters and conditions. A parameter which plays a
crucial role in the evolution of sex and recombination 
is epistatic 
interaction amongst gene loci \cite{Kouyos:2007}. Experiments have
shown that the 
individual locus do not always contribute independently to the fitness
of the whole sequence \cite{Wolf:2000,Phillips:2008} and the deviation
of the fitness from the independent loci model is a measure of the
epistatic interactions. The nature of epistasis is important in 
determining whether a mode of reproduction may be viable. 
For instance, 
in the absence of back mutations and 
recombination, a finite asexual population evolving on a nonepistatic
fitness landscape accumulates deleterious mutations irreversibly
(Muller's ratchet) 
\cite{Muller:1964,Felsenstein:1974}. 
But the degeneration can be effectively halted if 
synergistic epistasis is present
\cite{Kondrashov:1994,Jain:2008b}. On complex multipeaked
fitness landscapes that incorporate sign epistasis
\cite{Weinreich:2005b}, the effect of sex 
has been seen to depend on the detailed topography of the fitness landscape 
\cite{Kondrashov:2001,Visser:2009}.  

The epistatic interactions play an important
role in infinitely large populations as well.  
In a two locus model with the four possible
sequences denoted by $ab$, $Ab$, $aB$ and $AB$ and with respective
fitnesses $w_1, w_2, w_2, w_4 (> w_1, w_2)$, 
recombination reduces the frequency of
the favorable mutant $AB$ when epistasis parameter $e=w_4
w_1-w_2^2$ is positive but increases the $AB$ frequency for negative
$e$ \cite{Eshel:1970}. In this article, we ask: in an infinitely large
recombining population if all the population is initially
located at the sequence $ab$, how much time
$T$ does it take to get fixed to the double mutant $AB$ with fitness $w_4
> w_1$? The fixation time $T$ is expected to 
decrease with recombination for negative epistasis and increase for
positive epistasis \cite{Otto:1994,Feldman:1996}. These qualitative
trends are understandable from the results of \cite{Eshel:1970}: for $e < 0$,
as recombination acts in favor of the double mutant, it will get fixed
faster than in the asexual case while the reverse holds for $e > 0$
case. 

The main purpose of this article is to find 
analytical expressions for the fixation time $T$. To this end, we
develop a new method to handle the inherently nonlinear equations obeyed
by the genotype frequencies in the presence of recombination (see
Section~\ref{Model}). The basic idea of our approach is that at any
instant, only one of the genotypes dominate so that the equations 
can be expanded perturbatively in powers of the ratio of the
non-dominant genotype frequency to the dominant one. 

The rest of the article is organised as follows. We first define the
model under consideration in Section~\ref{Model}. The dynamics of the
population frequencies for various choices of epistasis are discussed in 
Section~\ref{timeevol}. The fixation time defined as the time at which
the double mutant frequency reaches a given finite fraction is 
calculated in  Section~\ref{Time}. The effect of initial
conditions on fixation time is considered in Section~\ref{asymm}. 
The last section discusses our
results which are summarised in Table \ref{summary}. 
 
%==========================================================================
%MODEL
%==========================================================================
\section{Model}
\label{Model}

We consider a two locus model with 
sequences denoted by $ab$, $Ab$, $aB$ and $AB$ and respective
fitnesses $w_1, w_2, w_3, w_4 (> w_1, w_2, w_3)$. The population at these
sequences evolves according to  mutation, selection and recombination
dynamics. In such models, several schemes
have been used to implement these basic processes such as 
recombination followed by mutation and then 
selection  \cite{Feldman:1996}, selection,
mutation and then recombination \cite{Kouyos:2006} and 
selection, mutation and recombination appearing as additive terms in
continuous time models \cite{Burger:1989}.  Here we work with a discrete
time, two locus model in which the mutation occurs after recombination 
and selection \cite{Stephan:1996}. The mutation
probability from $a$ to $A$ and $b$ to $B$ is given by $\mu$ but the
back mutations are neglected. The recombination between $ab$ and $AB$
or $aB$ and $Ab$ occurs with a
probability $r$. Denoting the population fraction at sequences 
$ab$, $Ab$, $aB$ and $AB$ at time $t$ by $x_1,...,x_4$ respectively
and time $t+1$ by $x_1',...,x_4'$, the time evolution occurs according
to the following nonlinear coupled equations: 
\bea
x_1' &=& \frac{ (1-\mu)^2 (w_1x_1-r D)}{{\bar w}(t)} \\
x_2' &=& \frac{ \mu (1-\mu) (w_1 x_1-r D)+  (1-\mu)(w_2 x_2+r D) }{{\bar w}(t)} \\
x_3' &=& \frac{ \mu (1-\mu) (w_1 x_1-r D)+  (1-\mu)(w_3 x_3+r D) }{{\bar w}(t)} \\
x_4' &=& \frac{ \mu^2 (w_1 x_1-r D)+   \mu (w_2 x_2+r D)+ \mu (w_3
  x_3+r D)+(w_4 x_4-r D) }{{\bar w}(t)}
\eea
Here $D(t)=(w_1 w_4 x_1(t) x_4(t)-w_2 w_3 x_2(t) x_3(t))/{\bar w}^2(t)$ is the linkage disequilibrium
at time $t$ and ${\bar w}(t)=\sum_{k=1}^4 w_k x_k(t)$ is  the average
fitness of the population.

In the following, we will work with $w_1=1, w_2=w_3$ and $w_4 > w_1,
w_2$ and initial condition $x_k(0)=\delta_{k,1}$. As a consequence,
$x_2(t)=x_3(t)$ for all $t > 0$. We define the epistasis parameter
$e=w_1 w_4 -w_2^2$ and will discuss four separate cases: (i) 
 zero epistasis which requires $w_2 > 1$ as $w_4 >
1$ (ii) negative epistasis (iii) 
positive epistasis and $w_2 > 1$  (iv) positive epistasis and $w_2 <
1$ (compensatory mutation). 
It is useful to write $x_k(t)=z_k(t)/\sum_{j=1}^4 z_j(t)$ 
where the $z_k$'s satisfy the following condition:
\be
\sum_{i=1}^4 z_i' = \sum_{i=1}^4 w_i z_i
\l{cond}
\ee
The unnormalised populations $z_k$'s obey the following set of equations,
\bea
z_1' &=& (1-\mu)^2 \left( z_1-r {\tilde D}\right) \l{z1eqn}\\
z_2' &=&  \mu (1-\mu) \left( z_1-r {\tilde D}\right)+  (1-\mu) \left(w_2 z_2+r
{\tilde D} \right) \l{z2eqn}\\
z_4' &=&  \mu^2 \left( z_1-r {\tilde D}\right)+  2 \mu \left(w_2 z_2+r {\tilde D} \right)+ \left(w_4 z_4-r  {\tilde D}\right)
\l{z4eqn}
\eea
with the initial condition $z_k(0)=\delta_{k,1}$. In the above
equations,
\be
{\tilde D}= D \sum_{i=1}^4 z_i= \frac{w_4 z_1 z_4-w_2^2
  z_2^2}{(\sum_{i=1}^4 w_i z_i)^2} \sum_{i=1}^4 z_i
\ee

\begin{figure}
\includegraphics[width=0.6 \linewidth,angle=270]{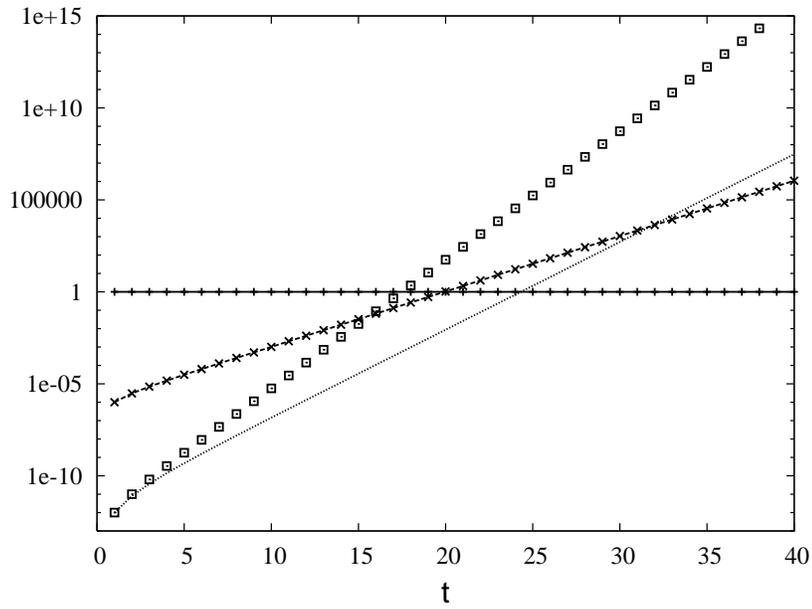}
\caption{Recombination probability $r=0$: Time evolution of $z_1$ (solid),
  $z_2$ (broken) and $z_4$ 
  (dotted) for $w_2=2, w_4=3, \mu=10^{-6}$, and $z_1$ (+),
  $z_2$ ($\times$) and $z_4$ 
  ($\boxdot$) for $w_2=2, w_4=5, \mu=10^{-6}$ using exact
  equations (\ref{z1eqn})-(\ref{z4eqn}). }  
\l{r0frac}
\end{figure}

For $r=0$, the above model reduces to the standard
quasispecies model for asexuals and can be solved exactly as the
population $z_i$'s obey linear equations 
\cite{Jain:2007b}.  For $\mu \to 0$, we find  
\bea
z_1(t) &\approx& 1 \\
z_2(t)  &\approx& \mu \left[\frac{w_2^t-1}{w_2-1} \right]\\
z_4(t)  &\approx& \mu^2 \left[ \frac{2 w_2}{w_2-1}
  \frac{w_4^t-w_2^t}{w_4-w_2}-\frac{w_2+1}{w_2-1} \frac{w_4^t-1}{w_4-1} \right]
  \l{z4r0}
\eea
Figure \ref{r0frac} shows the time evolution of the populations
$z_k$'s  when $r=0$  for negative and positive epistasis. 

With nonzero recombination, it does 
not seem possible to solve the above equations for $z_i(t)$
exactly due to the bilinear terms in linkage disequilbrium
$D$. However, in the next section, we will obtain approximate
expressions for the unnormalised population $z_i$. 
%for weak selection i.e. $s_2, s_4 \ll 1$ where $w_2=1+s_2, w_4=1+s_4$. 

%==========================================================================
%POPULATION
%==========================================================================
\section{Time evolution of populations}
\l{timeevol}

As we shall see, the dynamics of
population $z_i$'s can be divided in following three dynamical phases:
(i) $z_1 \gg z_2, z_4$ (phase I) (ii) $z_2 \gg z_1, z_4$ (phase
II) and (iii) $z_4 \gg z_1, z_2$ (phase III). Thus we can expand
equations (\ref{z1eqn})-(\ref{z4eqn}) in powers of $z_2/z_1, z_4/z_1$
in phase I, $z_1/z_2, z_4/z_2$ in phase II and similarly, $z_1/z_4,
z_2/z_4$ in phase III.  The time scale at which a
phase ends is obtained by matching the solutions of the relevant
populations in the two phases. In the crossover region however the
above assumptions are not expected to hold strictly. But as we shall see, the
fixation time is nevertheless well approximated. 
Note that the perturbation expansions here are different from a 
small $r$ expansion \cite{Kouyos:2006}.

%==========================================================================
%NONEPISTATIC
%==========================================================================
\subsection{No epistasis}

When $w_4=w_2^2$, the epistasis parameter $e=0$. The dynamics of 
populations $x_i$'s and $z_i$'s evolving for such a fitness choice is shown in
Fig.~\ref{nullfrac}. 
%It is straightforward to check that the linkage disequilibrium $D(t)$ obeys the following evolutionequation for any $e$:\beD'= \left(\frac{\mu \sum_i z_i}{\sum_i w_i z_i} \right)^2 \left[  \left(w_4 -r \frac{\sum_i w_i z_i}{\sum_i z_i} \right) D + \frac{e    z_2^2}{(\sum_i z_i)^2}\right]\l{Dexact}\ee
For $e=0$, the linkage disequilibrium ${\tilde D}(t)$ obeys the following
evolution equation:
\bea
{D}' &=&  \frac{w_2^2 (1-\mu)^2 \left[w_2^2 (z_1 z_4-z_2^2)-r D S_1 S_2
  \right]}{\left[(1-\mu+\mu w_2)^2 z_1+2 w_2^2
  (1-\mu+\mu w_2) z_2+w_2^2 z_4-r (1-\mu)^2 (1-w_2)^2 D S_1 \right]^2} \no
\\
&\propto& w_2^2 (z_1 z_4-z_2^2)-r D S_1 S_2
\eea
where $S_1=\sum_{i=1}^4 z_i$ and $S_2=\sum_{i=1}^4 w_i z_i$. As
$D \propto z_1 z_4-z_2^2$ for $e=0$ , it follows that $D' \sim
D$. Therefore, 
%\be
%z_1' z_4'-z_2'^{2} = (1-\mu)^2 (z_1 z_4-z_2^2) \left(w_4- r
%\frac{\sum_i w_i z_i}{\sum_i z_i} \right)
%\ee
%Since $z_1 z_4-z_2^2$ at $t=0$ is zero, it follows from the aboveequation that $z_1 z_4-z_2^2=0$ for all $t > 0$. In other words, 
if the population has zero linkage disequilibrium to
start with, it remains zero for all times in the absence of epistasis
\cite{Eshel:1970}. 
Thus for $e=0$, the population $z_k$'s obey following
linear equations, 
\bea
z_1' &=& (1-\mu)^2 z_1 \no \\
z_2' &=&  \mu (1-\mu) z_1 +  (1-\mu) w_2 z_2\no \\
z_4' &=&  \mu^2 z_1 +2 \mu w_2 z_2+ w_4 z_4  \no
\eea
which can be easily solved. For small $\mu$, we obtain
\bea
z_1 (t) &\approx& 1 \l{z1e0} \\
z_2 (t) &\approx& \mu \left(\frac{w_2^t-1}{w_2-1} \right) \l{z2e0} \\
z_4 (t) &\approx&  \left[ \mu \left(\frac{w_2^t-1}{w_2-1} \right) \right]^2~.
\l{z4e0} \eea
\begin{figure}
\includegraphics[width=0.6 \linewidth,angle=270]{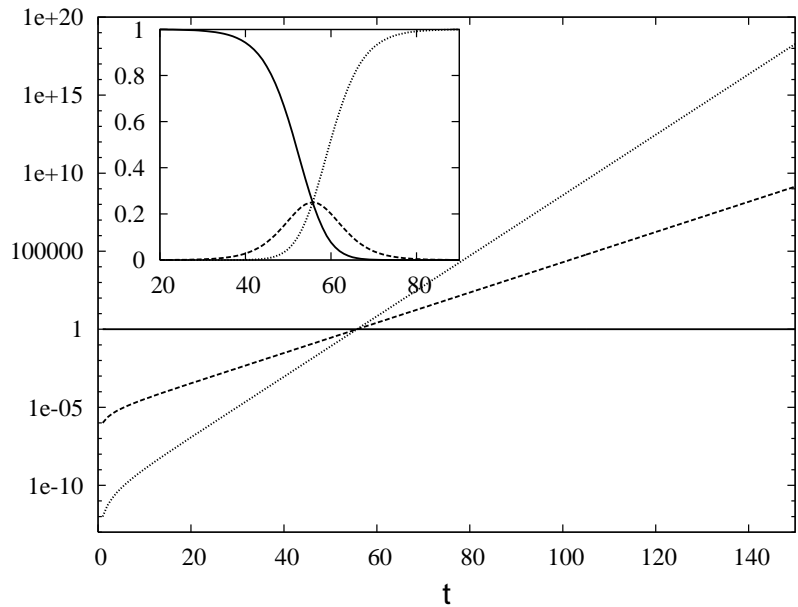}
\caption{Non-epistatic interaction: Time evolution of $z_1$ (solid),
  $z_2$ (broken) and $z_4$ 
  (dotted) for $w_2=1.25, w_4=1.5625, r=0.1, \mu=10^{-6}$ using exact
  equations (\ref{z1eqn})-(\ref{z4eqn}). The normalised fractions are shown in the inset.}
\l{nullfrac}
\end{figure}

From the above solution, we see that at time $\tau_1$ at which
phase I ends, $z_1 
z_4-z_2^2=z_4-z_2^2=0$ so that $z_4=z_2$ and therefore the phase II is
absent in this case. 

%==========================================================================
%NEGATIVE
%==========================================================================
\subsection{Negative epistasis}
\l{neg}

We now consider the case when epistasis is negative, 
$w_4 < w_2^2$. The time evolution of 
populations for this fitness scheme is shown in Fig.~\ref{r0frac} for
$r=0$ and Fig.~\ref{negfrac} for $r > 0$. In both cases, the
population $z_1$ dominates at short times 
followed by $z_2$ and finally $z_4$ takes over. This behavior is also
reflected in the normalised populations $x_k$'s shown in the inset of 
Fig.~\ref{negfrac}. 

\begin{figure}
\includegraphics[width=0.6 \linewidth,angle=270]{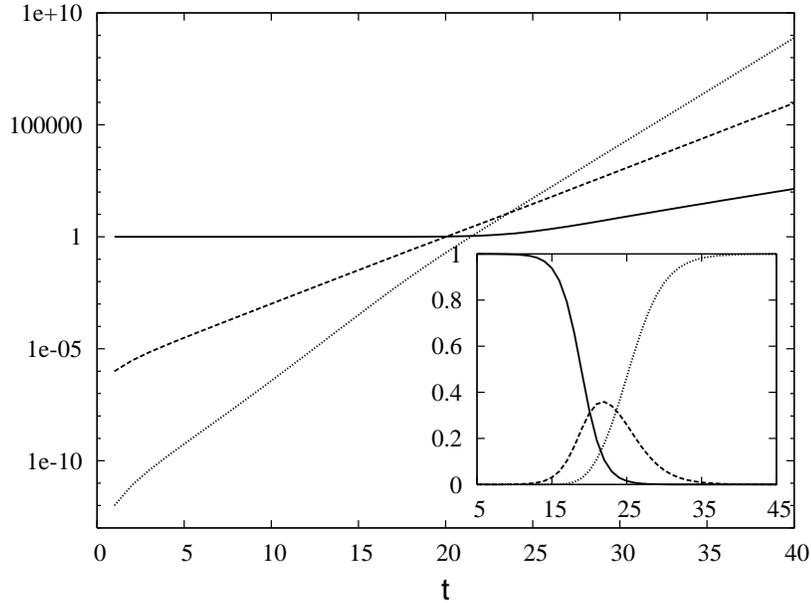}
\caption{Negative epistasis: Time evolution of $z_1$ (solid),
  $z_2$ (broken) and $z_4$ 
  (dotted) for $w_2=2, w_4=3, r=0.1, \mu=10^{-6}$ using exact
  equations (\ref{z1eqn})-(\ref{z4eqn}). }
\l{negfrac}
\end{figure}

%---------------------------------------------------------------------------

\noindent{\underline{\it Phase I.}} 
Since $z_{k}(0)=\delta_{k,1}$, initially $z_2, z_4 \ll z_1$ so that the
sum $\sum_{i=1}^4 z_i \approx z_1$. Using this in the expression for
${\tilde D}$, we find that ${\tilde D} \approx w_4
z_4-(w_2^2 z_2^2/z_1)$. Taking $\mu \to 0$ in (\ref{z1eqn})-(\ref{z4eqn}),
we then obtain 
\bea
z_1'&\approx& 
z_1+r \left(\frac{w_2^2 z_2^2-w_4 z_4 z_1}{z_1} \right)\no \\
z_2'
&\approx &\mu z_1+w_2 z_2- r \left(\frac{w_2^2 z_2^2-w_4 z_4 z_1}{z_1} \right) \no \\
z_4'&\approx& 
\mu^2 z_1+2 \mu w_2 z_2+ w_4 z_4+ r \left(\frac{w_2^2 z_2^2-w_4 z_4 z_1}{z_1} \right)\no 
\eea
Since $z_2/z_1, z_4/z_1 \ll 1$, we can write $z_1' \approx z_1$
which immediately gives $z_1(t)=1$ for $t < \tau_1$ where $\tau_1$ is
the time at which phase I ends. In the equation for $z_2$ ($z_4$), the first
term on the RHS is the mutation term which signifies that 
even if $w_2$ ($w_4$) and $r$ are equal to
zero, the population $z_2$ ($z_4$) will remain nonzero due to a
constant mutational supply from $ab$. The second term on the RHS of
the $z_2$ equation is the selection term and
the last two terms are due to recombination. It turns out that the
recombination term can be ignored to yield $z_2'  \approx  \mu z_1+w_2 z_2$.  
This approximation is later justified by showing that indeed the
last two terms are negligible compared to $w_2 z_2$. In summary, we have
\bea
z_1' &\approx  &z_1 \\
z_2' & \approx & \mu z_1+w_2 z_2 
\l{z2P1}\\
z_4'& \approx & \mu^2 z_1 +2 \mu w_2 z_2 + (1-r) w_4 z_4+ r w_2^2 \frac{z_2^2}{z_1}
\eea
On solving the above equations, we obtain
\bea
z_1 (t) & \approx & 1 
\l{z1P1neg} \\
z_2 (t) & \approx & \mu \left( \frac{w_2^t-1}{w_2-1} \right) \l{z2P1neg}\\
z_4 (t) & \approx & \mu^2 \left[ \frac{(w_4 (1-r))^t-1}{w_4 (1-r)-1}\right]+\frac{2
  \mu^2 w_2}{w_2-1} \left[
  \frac{(w_4 (1-r))^t-w_2^t}{w_4 (1-r)-w_2}-\frac{(w_4 (1-r))^t-1}{w_4 (1-r)-1}
  \right] \no \\
&+& \frac{r \mu^2 w_2^2}{(w_2-1)^2} \left[ \frac{(w_4 (1-r))^t-w_2^{2
      t}}{w_4 (1-r)-w_2^2} -2
  \frac{(w_4 (1-r))^t-w_2^t}{w_4 (1-r)-w_2}+\frac{(w_4 (1-r))^t-1}{w_4 (1-r)-1}\right] \l{z4P1neg}
\eea
We check that the $r=0$ limit is recovered from the above solution. 
From the last equation, we find that for $r > 0$, the growth rate of population
$z_4$ is given by $\max\{w_4 (1-r), w_2^2\}$. 
Since $r$ must be positive, for $e
< 0$, the growth rate of $z_4$ is $w_2^2$ and we have 
\be
z_4(t) \approx \frac{\mu^2 r w_2^2}{(r w_4+|e|) (w_2-1)^2} \times w_2^{2 t}~,~r
> 0
\l{Az4P1neg}
\ee
Using the above solutions, it can be checked  that $w_2 z_2 \gg r
(w_2^2 z_2^2-w_4 z_4)$ thus justifying (\ref{z2P1}). 

It is evident from (\ref{z2P1neg}) and (\ref{Az4P1neg}) that when $z_2$
becomes one, $z_4=r w_2^2/(r w_4+|e|) < 1$ so that $z_2$ intersects $z_1$ 
before $z_4$. The time $\tau_1$ at which $z_2(\tau_1)=1$ is given by 
\be
\tau_1= \frac{1}{\ln w_2} \ln \left( \frac{w_2-1}{\mu} \right)
\l{tau1neg}
\ee
which is independent of $r$. 

%---------------------------------------------------------------------------
\noindent{\underline{\it Phase II.}} After time $\tau_1$, the
population $z_2 \gg z_1, z_4$ and therefore the sum $\sum_{i=1}^4 z_i
\approx 2 z_2$. For weak selection, this gives ${\tilde D} \approx
-z_2/2$. Thus in this phase, $z_k$'s obey the following equations:
\bea
z_2' &\approx&  \left( w_2-\frac{r}{2} \right) z_2\\
z_4' &\approx& w_4 z_4+ \frac{r}{2} z_2  \\
z_1' &\approx& z_1+ \frac{r}{2} z_2
\eea
As the equation for $z_2$ is decoupled from $z_1$ and $z_4$, we can 
first solve for $z_2$ and then use the solution to find $z_1$ and
$z_4$. This finally gives 
\bea
z_2(t) &\approx& \left( w_2-\frac{r}{2} \right)^{t-\tau_1} z_2(\tau_1)
\l{z2P2neg} \\
z_4(t)  &\approx&  w_4^{t-\tau_1} z_4(\tau_1) +\frac{r}{2} \frac{\left[w_2-(r/2) \right]^{t-\tau_1}-w_4^{t-\tau_1}
}{w_2-(r/2)-w_4}  z_2 (\tau_1)  \l{z4P2neg} \\
z_1(t) &\approx& z_1(\tau_1)+ \frac{r}{2} \frac{\left[w_2-(r/2) \right]^{t-\tau_1}-1
}{w_2-(r/2)-1} z_2(\tau_1) 
\l{z1P2neg}
\eea
where $z_1(\tau_1)=z_2(\tau_1)=1,
z_4(\tau_1)=r w_2^2/(r w_4+|e|)$. 

The time $\tau_2$ at which $z_4$ overtakes $z_2$ is given by 
\be
\tau_2 = \tau_1+ \frac{1}{\ln\left[2 w_4/(2 w_2-r)\right]} \ln\left[
  \frac{2 (r w_4+|e|) (w_2-w_4-r)}{ \left[2(w_2-w_4)-r \right] r w_2^2-r^2 w_4-r |e|}\right]~,~r > 0
\l{tau2}
\ee
We check that for $e=0$, the time $\tau_2-\tau_1$ during which phase
II is present vanishes. To understand how $\tau_2$ varies with $r$, 
consider the $r$-dependent term $g(r)$ in $\tau_2$ which can be
rewritten as 
\bea
g(r) &=&\frac{1}{\ln\left[2 w_4/(2 w_2-r)\right]} \ln \left[1 +
  \frac{|e|}{r} \frac{(1-r) (r+2 (w_4-w_2))}{|e|+r w_4-w_2^2 
\left[2 (w_2-w_4)-r \right]} \right] \no \\
&\approx& \frac{1}{\ln(w_4/w_2)} \ln \left[1 + \frac{|e|}{r} \frac{(1-r) (r+2 (w_4-w_2))}{|e|+r w_4-w_2^2 
\left[2 (w_2-w_4)-r \right]} \right] \no \\
&=& \frac{1}{\ln(w_4/w_2)} \ln \left[1 + {\cal R} \right] \no
\eea
The ratio ${\cal R}=1$ when $r$ satisfies the quadratic equation
$w_2^2 r^2+(w_4-w_2) (2 w_2^2-w_4)r-(w_4-w_2) |e|=(r-r_+) (r+|r_-|)=0$ where $r_+ (r_-)$ is
the positive (negative) root of the quadratic equation. For $r \ll r_+$,
the ratio ${\cal R} \gg 1$ so that $\ln (1+{\cal R}) \approx \ln {\cal R}$. For $r \gg r_+, {\cal R}
\ll 1$ and $\ln (1+{\cal R}) \approx {\cal R}$. Using these approximations in
the expression for $g(r)$ above, we find that
\be
g(r) \sim \begin{cases} \ln\left[2 (w_4-w_2)/r \right] &~,~  r \ll r_+ \\
             |e|/r
&~,~ r \gg r_+
\end{cases}
\ee
Thus the time $\tau_2-\tau_1$ decreases slowly as $\ln (1/r)$ for small $r$
and as $1/r$ for large $r$. Due to these properties of $\tau_2$, the
single mutant population can dominate for appreciable time 
interval for small $r$.

%---------------------------------------------------------------------------
\noindent{\underline{\it Phase III.}}
For $t > \tau_2$, the
population $z_4 \gg z_1, z_2$ so that $\sum_i z_i \approx z_4$. Due to
this, ${\tilde D} \approx (z_1/w_4)- ((w_2^2 z_2^2)/(w_4^2 z_4))$. For 
small $\mu$, the equation for $z_k$'s in (\ref{z1eqn})-(\ref{z4eqn})
can thus be simplified to give 
\bea
z_4' &\approx& w_4 z_4 \l{z4P3} \\
z_2' &\approx& w_2 z_2 \l{z2P3} \\
z_1' &\approx& \left(1-\frac{r}{w_4} \right) z_1 + \frac{r
  w_2^2}{w_4^2} \frac{z_2^2}{z_4} \l{z1P3} 
\eea
where we have neglected the recombination term contribution to the
equation for $z_2$ by assuming $w_2 z_2 \gg r z_1/w_4$
(see below). From the first two equations, we see that $z_2 \sim
w_2^t$ and $z_4 
\sim w_4^t$. Thus the last term in the equation for $z_1$ can
contribute when $e < 0$. Explicitly, we obtain
\bea
z_4(t) &=& w_4^{t-\tau_2} z_4(\tau_2) \l{z4P3neg}\\
z_2(t) &=& w_2^{t-\tau_2} z_2(\tau_2) \l{z2P3neg}\\
z_1(t) &=& (1-\frac{r}{w_4})^{t-\tau_2} z_1(\tau_2)+ r
\frac{w_2^2}{w_4^2}\frac{z_2^2(\tau_2)}{z_4(\tau_2)}
~\frac{(1-(r/w_4))^{t-\tau_2}-(w_2^2/w_4)^{t-\tau_2}}{(1-(r/w_4))-(w_2^2/w_4)} \l{z1P3neg}
\\
&\approx& 
\frac{r w_2^2 z_2^2(\tau_2)}{w_4 (r+|e|) z_4(\tau_2)} \left( \frac{w_2^2}{w_4}
\right)^{t-\tau_2}~,~e < 0
\eea
where $z_k(\tau_2)$ are given by (\ref{z2P2neg})-(\ref{z1P2neg}) at time $t=\tau_2$. 
From the above solution, it is easily verified that $w_2 z_2 \gg r
z_1/w_4$ is a good approximation for $t > \tau_2$. 

%==========================================================================
%POSITIVE
%==========================================================================
\subsection{Positive epistasis}

We now turn to the case when  epistasis is positive. 
The condition $w_4 > w_2^2$ can be satisfied for $w_2 < 1$ and $w_2 >
1$. For $w_2 > 1$, the time evolution of  
populations for this fitness scheme is shown in Fig.~\ref{posfrac} for
$e < r w_4$ and $e > r w_4$. The reason for this distinction will be explained
below. The dynamics of the populations $z_1, z_2$ and $z_4$ for $w_2 <
1$ are shown in Fig.~\ref{compfrac}. Note that phase II is absent in
all these cases. 

\noindent{\underline{\it Phase I.}} As discussed for $e < 0$, in this phase, $z_1 \gg z_2, z_4$ and can be well approximated by one, $z_1(t) \approx 1$ for $t < \tau_1$. Then the populations $z_2$ and $z_4$ obey the following equations: 
\bea
z_2' &\approx& \mu + w_2 z_2 +r (w_4 z_4-w_2^2 z_2^2) \l{z2P1pos} \\
z_4' &\approx& \mu^2  +2 \mu w_2 z_2 + w_4 (1-r) z_4 +r w_2^2 z_2^2 \l{z4P1pos}
\eea
Figures \ref{posfrac} and \ref{compfrac} show that initially $z_2 >
z_4$ but 
after some time ($< \tau_1$), $z_4$ can overtake $z_2$ while both
$z_2, z_4 < 1$. This behavior is characteristic of positive epistasis
as can be seen from Fig.~\ref{r0frac} for $r=0$ also. For this reason,
the population $z_2$ can get a contribution from $z_4$ in phase I when
$e > 0$ and we need to retain the $r$-dependent term in the equation
for $z_2$. The equation for $z_4$ remains the same as for negative
epistasis. 
Since $z_4$ appears linearly in the above equations for $z_2$ and
$z_4$, it is possible to eliminate $z_4$ from the equation for $z_2$ 
and express it in terms of $z_2$ alone. This gives a three term
recursion relation for $z_2$:   
\bea
z_2(t+1) &=& (1+r w_4-w_4) \mu+r w_4\mu^2+ (2 \mu
r w_4-w_4+r w_4) w_2 z_2(t-1) \no \\
&+& (w_2+w_4-r w_4) z_2(t)+r w_4 w_2^2
z_2^2(t-1)-r w_2^2 z_2^2(t)~,~t \ge 1
\l{z2P1pos2}
\eea
with initial conditions $z_2(0)=0$ and $z_2(1)=\mu$. We will find the
solution to the nonlinear equation for $z_2(t)$ iteratively
\cite{Bender:1999}.  We first find the
solution $f_0(t)$ of the above difference equation for $z_2$ when the nonlinear
terms are set to zero. The corrections to $z_2(t)$ arising due to
nonlinearity will then be determined by writing $z_2(t)=f_0(t)
(1+f_1(t))$. 

The solution $f_0(t)$ satisfies the following linear, inhomogeneous
difference equation:
\be
f_0(t+1)=B_0 f_0(t)+ C_0 f_0(t-1) +A_0~,~t \ge 1
\ee
where $A_0=(1+r w_4-w_4) \mu+r w_4\mu^2, B_0=(w_2+w_4-r w_4), C_0=(2 \mu
r w_4-w_4+r w_4) w_2$. 
The solution of this linear equation subject to $f_0(0)=0, f_1(1)=\mu$
can be found by using the method of variation of parameters
\cite{Bender:1999} and  is given by 
\be
f_0(t) = \frac{\mu (1-\alpha_+)-A_0}{(\alpha_+-\alpha_-) (1-\alpha_+)}
\alpha_+^t+\frac{\mu (1-\alpha_-)-A_0}{(\alpha_--\alpha_+) (1-\alpha_-)}
\alpha_-^t+\frac{A_0}{(1-\alpha_-)(1-\alpha_+)} \no 
\ee
where
\be
\alpha_{\pm}= \frac{(w_4-r w_4+w_2) \pm \sqrt{(w_4-r w_4-w_2)^2+8 \mu r w_4 w_2
}}{2} \no
\ee
For our purposes, it is sufficient to retain terms to ${\cal
  O}(\mu^2)$ in  the last expression which gives 
\bea
f_0(t)&\approx& \left(\mu +
\frac{\mu^2 r w_4\left[2 w_2(1-w_2)+(1+w_2) (w_4-w_2-r w_4) \right]}{(w_2-1)
  (w_4-w_2-r w_4)^2} \right)\left[\frac{w_2^t-1}{w_2-1} \right] \no \\
&+& \frac{\mu^2 r w_4
  (w_4+w_2-r w_4)}{(w_4-r w_4-w_2)^2} \left[\frac{(w_4-r
    w_4)^t-1}{w_4-r w_4-1}\right]
  \l{f0t} 
\eea
Note that the above solution consists of two growth rates
for $z_2$ namely $w_2$ and $w_4 (1-r)$. 

Using $z_2(t) \approx f_0(t)$ in (\ref{z4P1pos}) and keeping terms to
${\cal O}(\mu^2)$, we get (\ref{z4P1neg}) for $z_4(t)$. It follows
that the population $z_4$ does not grow if $r > r_c=(w_4-1)/w_4$ for 
$w_2 < 1$. But for $w_2 > 1$, the population $z_4$ always
grows and the growth rate is given by $\max\{w_4 (1-r), w_2^2\}$ as in the
case of negative $e$. In the following, we will discuss the two cases
$w_2 > 1$ and $w_2 < 1$ separately.  

%^^^^^^^^^^^^^^^^^^^^^^^^^^^^^^^^^^^^^^^^^^^^^^^^^^^^^^^^^^^^^^^^^^^^^^^
%w2 > 1
%^^^^^^^^^^^^^^^^^^^^^^^^^^^^^^^^^^^^^^^^^^^^^^^^^^^^^^^^^^^^^^^^^^^^^^^
\begin{figure}
\includegraphics[width=0.42 \linewidth,angle=270]{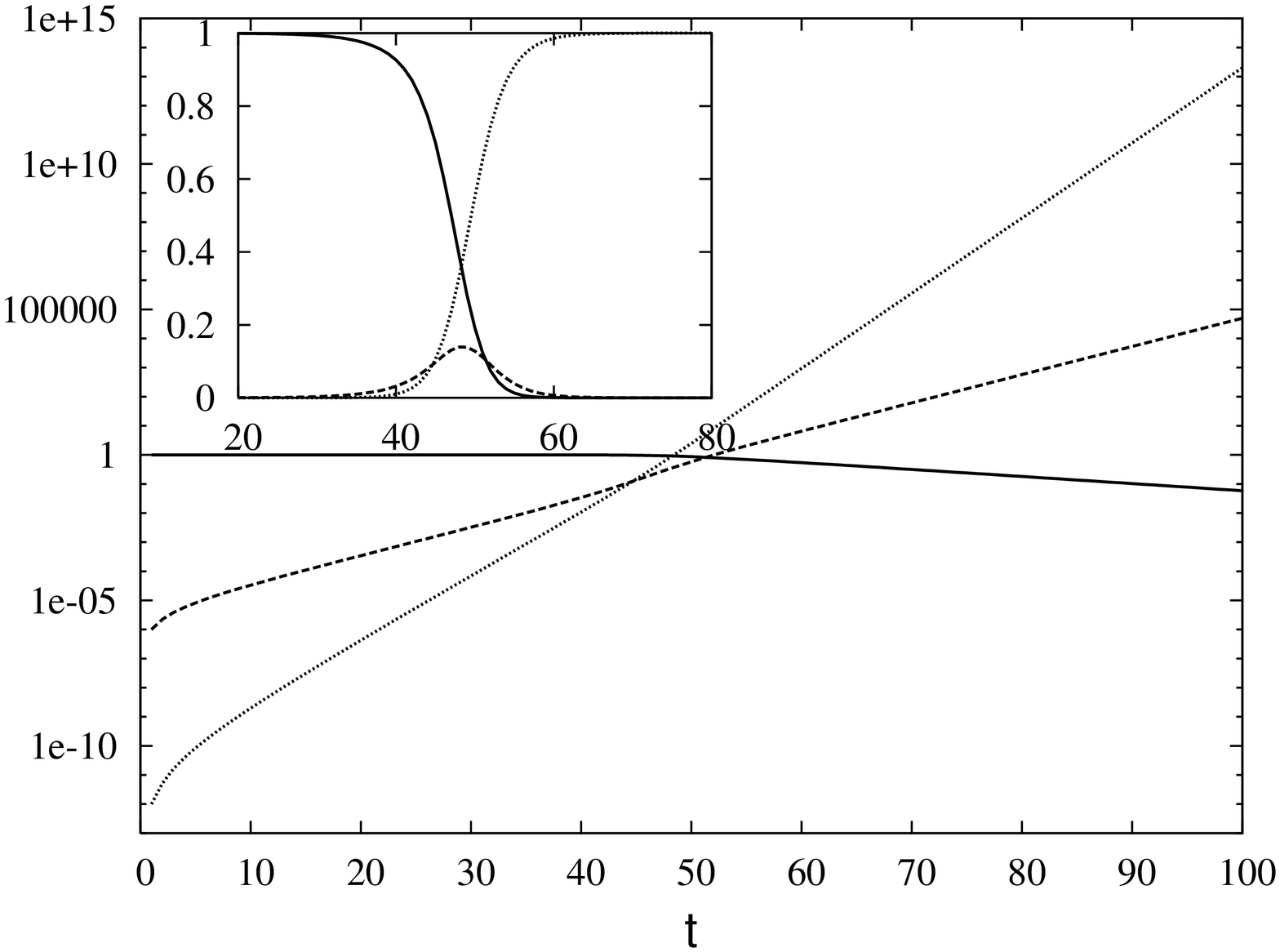}
\includegraphics[width=0.42 \linewidth,angle=270]{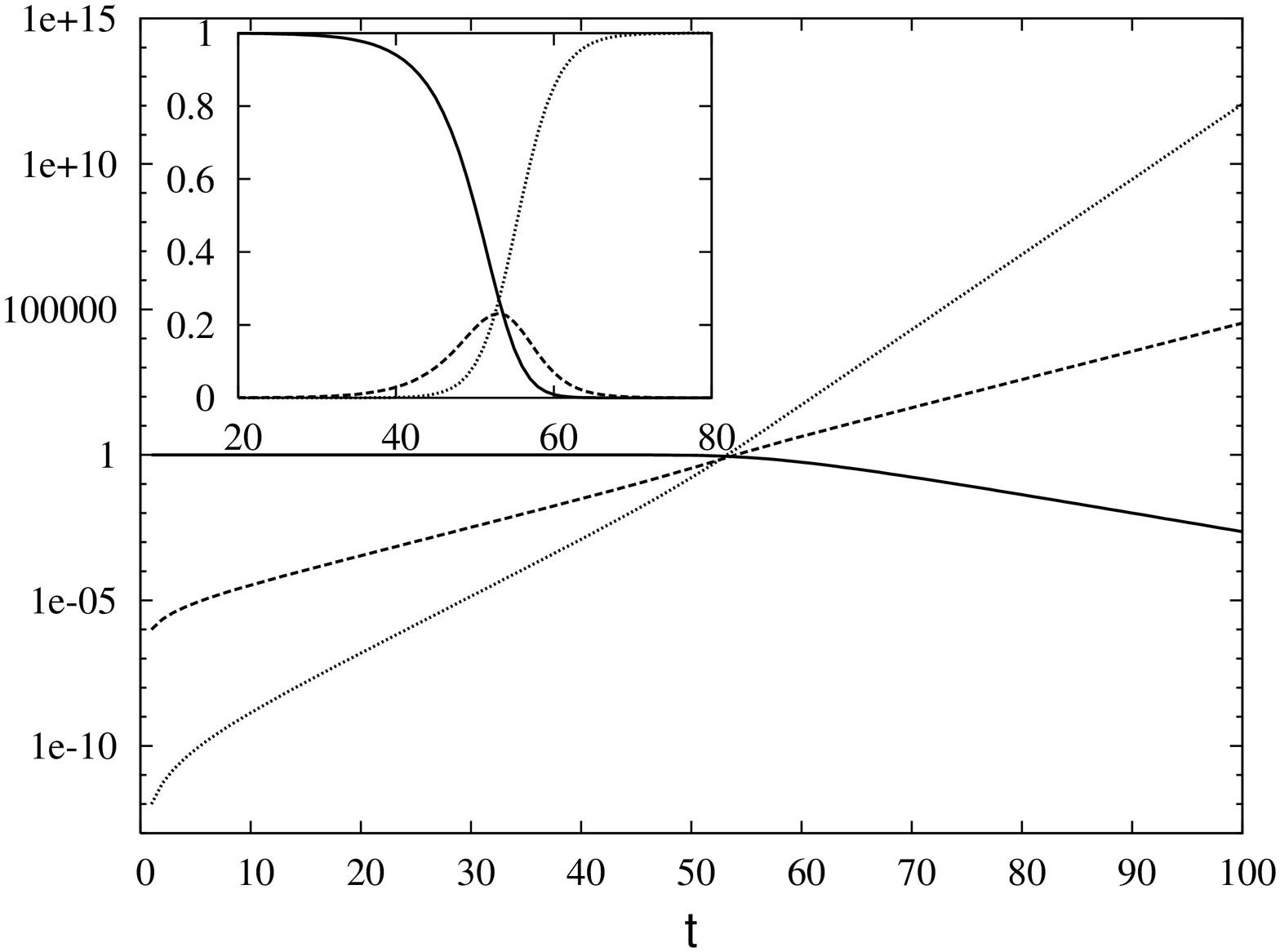}
\caption{Positive epistasis: Time evolution of $z_1$ (solid),
  $z_2$ (broken) and $z_4$ 
  (dotted) for (a) $w_2=1.25, w_4=1.8125, r=0.1, \mu=10^{-6}$ (b) $w_2=1.25,
  w_4=1.8125, r=0.4, \mu=10^{-6}$ using exact
  equations (\ref{z1eqn})-(\ref{z4eqn}). The normalised fractions are shown in the inset.}
\l{posfrac}
\end{figure}

\noindent \underline{$w_2 > 1$:} When $w_2 > 1$, due to
(\ref{z4P1neg}), the following subcases arise for $z_4(t)$:   
\be
z_4(t) \approx 
\begin{cases} 
\frac{ \mu^2 e (1-r) (w_4-r w_4+w_2)}{(e-r w_4) (w_4-r w_4-1) (w_4-r w_4-w_2)}\times (w_4-r w_4)^t~ &,~r< e/w_4 \\
\frac{\mu^2 r w_2^2}{(r w_4-e) (w_2-1)^2} \times w_2^{2 t}~ &,~ r > e/w_4
\end{cases}
\l{z4P1pos2}
\ee
For $r > e/w_4$, it follows from (\ref{z4P1pos2}) that when $z_2$
becomes one, $z_4=r w_2^2/(r w_4-e) > 1$ so that $z_4$ hits unity
before $z_2$ and thus phase II is absent. The time $\tau_1$ at which 
$z_4(\tau_1)=1$ is given by 
\be
\tau_1(r)=  \frac{1}{\ln w_2} \ln \left( \frac{w_2-1}{\mu w_2} \right)+
\frac{1}{2 \ln w_2} \ln \left(w_4-\frac{e}{r} \right)~,~ r > e/w_4
\l{tau1rgtre}
\ee
For $r \gg e$, the last term in the above expression (and hence
$\tau_1$) increases as $\sim -e/(r w_4)$ with increasing $r$.

For $r < e/w_4$, using (\ref{z4P1pos2}), we find that the time $\tau_1$ at
which $z_4(\tau_1)=1$ is given by 
\be
\tau_1 (r) = \frac{1}{\ln (w_4-r w_4)} \ln \left[ \frac{(e-r w_4)
    (w_4-1-r w_4)(w_4-w_2-r w_4)}{\mu^2 e (1-r) (w_4+w_2-r w_4) } \right] ~,~r < e/w_4 \\
\l{tau1lsse}
\ee
For $r=0$, we have 
\be
\tau_1(0) \approx \frac{1}{\ln w_4} \ln \left[ \frac{
    (w_4-1)(w_4-w_2)}{\mu^2 (w_4+w_2) } \right] 
\ee
which matches the one obtained using (\ref{z4r0}) or (\ref{z4P1neg})
for $w_2 \ll w_4$. To find the behavior of $\tau_1$ for $r \ll e/w_4$, we
rewrite the expression for $\tau_1(r)$ as 
\bea
&&\ln(w_4-r w_4) \tau_1(r) - \ln w_4 \tau_1(0) \no \\
&=& \ln \left[ \left(1- \frac{r w_2^2}{e (1-r)}\right)
  \left(1- \frac{r w_4}{w_4-1}\right) \left(1- \frac{r w_4}{w_4-w_2}\right)
  \left(1-\frac{r w_4}{w_4+w_2} \right)^{-1}  \right] \no
\eea
Using the inequality $r < r w_4 < e < w_4-w_2 < w_4 -1 < w_4+w_2$ in the last
equation, we find 
\be
\ln (w_4-r w_4) \tau_1(r) \approx \ln w_4 \tau_1(0)+ \ln
\left(1-\frac{r w_2^2}{e}\right)
\ee
The above expression can be further simplified to give
\be
\tau_1(r) \approx \tau_1(0) \left(1+\frac{r}{\ln w_4} \right)~,~r
\ll e/w_4
\l{tau1rle}
\ee
which shows that $\tau_1$ increases linearly with $r$ for $r \ll e/w_4$.

%^^^^^^^^^^^^^^^^^^^^^^^^^^^^^^^^^^^^^^^^^^^^^^^^^^^^^^^^^^^^^^^^^^^^^^^
%w2 < 1
%^^^^^^^^^^^^^^^^^^^^^^^^^^^^^^^^^^^^^^^^^^^^^^^^^^^^^^^^^^^^^^^^^^^^^^^
\begin{figure}
\includegraphics[width=0.6 \linewidth,angle=270]{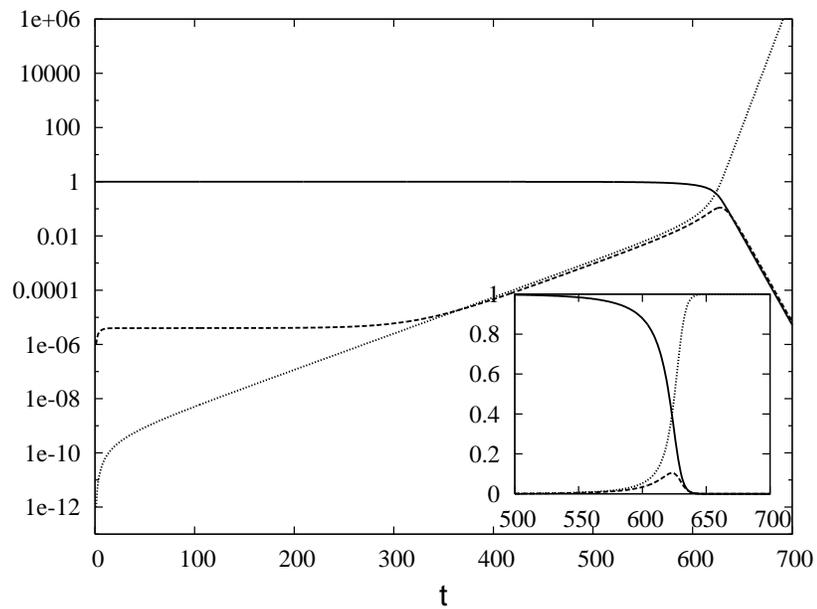}
\caption{Compensatory mutation: Time evolution of $z_1$ (solid), $z_2$
  (broken) and $z_4$ 
  (dotted) for $w_2=0.75, w_4=1.25, r=0.175, \mu=10^{-6}$ using exact
  equations (\ref{z1eqn})-(\ref{z4eqn}). The normalised fractions are
  shown in the  inset.}  
\l{compfrac}
\end{figure}

\noindent \underline{$w_2 < 1$:}  It is known that for an infinite
population, there exits a critical recombination fraction $r_c$ 
beyond which a 
population initially located at $ab$ cannot cross the intermediate
fitness valley and reach the double mutant
fitness peak \cite{Crow:1965,Eshel:1970,Higgs:1998,Park:2009}. For our
model, as discussed 
above, $z_4$ can grow (and hence $x_4$ can be fixed) provided $w_4 (1-
r) > 1$ or $r < r_c \approx (w_4-1)/w_4$.

Due to (\ref{f0t}) obtained by dropping nonlinear terms in the
equation for $z_2$, both $z_2$
and $z_4$ increase as $(w_4-r w_4)^t$. But Fig.~\ref{compfrac} shows that
$z_2$ and $z_4$ do not continue to grow in this manner but rise sharply  as the
end of phase I approaches. To understand this, it is essential to
include the nonlinear terms 
in the equations (\ref{z2P1pos}) and (\ref{z4P1pos}) for $z_2$ and
$z_4$.  
To this end, we write $z_2(t)=f_0(t) \left[1+f_1(t) \right]$ where
$f_0(t)$ given by (\ref{f0t}) reduces to 
\be
f_0(t) \approx \frac{\mu}{1-w_2}+ \frac{\mu^2 r w_4 (w_4+w_2-r
  w_4)}{(w_4-r w_4-w_2)^2} \left[\frac{(w_4-r w_4)^t}{w_4-r w_4-1}\right]
 \ee
for $w_2 < 1, w_4 (1-r) > 1$. 
 As we shall later, $f_1(t)$ remains close to zero for short times but
 contributes substantially at large times. Using this form of $z_2$ in
 (\ref{z2P1pos2}) and neglecting the quadratic terms in $f_1$, we find
 that $f_1(t)$ obeys the following approximate linear inhomogeneous
 equation with  time-dependent coefficients, 
\be
f_1(t+1)=B_1(t) f_1(t)+C_1(t) f_1(t-1)+A_1(t)
\l{f1teqn}
\ee
where the coefficient
\bea
B_1 (t)&=& w_2+w_4-r w_4-2 r w_2^2 f_0(t) \\
C_1 (t)&=& (2 \mu r w_2-w_4+r w_4) w_2+2 r w_4 w_2^2 f_0(t) \\
A_1 (t)&=& r (w_4-1) w_2^2 f_0(t)
\eea
For  $w_4 -r w_4 > 1$, we define $\epsilon=r_c-r$. For
$\epsilon \to 0$, $f_0(t) \approx a (1+\epsilon w_4)^t$
where 
\be
a=\frac{\mu^2 r_c (1+w_2)}{(1-w_2)^2 \epsilon}
\l{adef}
\ee
and the coefficients are given by 
\bea
B_1(t)  &\approx&  w_2+1-2 r_c w_2^2 f_0(t) \\
C_1(t)  &\approx& -w_2+2 r_c w_4 w_2^2 f_0(t) \\
A_1(t)  &\approx& r_c^2 w_4 w_2^2  f_0(t)
\eea

Writing the difference equation (\ref{f1teqn}) for $f_1(t)$ as a
differential equation, we get
\be
\frac{d f_1(t)}{dt}+ \frac{1-B_1(t)-C_1(t)}{1+C_1(t)} f_1(t)= \frac{A_1(t)}{1+C_1(t)}
\ee
with the initial condition $f_1(0)=0$.  It is straightforward to solve the above differential equation and we obtain
\be
f_1(t) \approx  \frac{1}{2} \left[  \left( b+ c (1+\epsilon w_4)^t
  \right)^{\alpha}-1   \right] 
\ee
where 
\bea
b &=& \frac{1-w_2}{1-w_2+2 r_c w_4 w_2^2 a} \\
c &=& 1-b \\
\alpha &=& \frac{r_c}{w_4 \epsilon}
\eea
The second term in the parentheses in the above equation can be
neglected for $t \ll t_1=\ln\left[(1-w_2)/(2 r_c w_4 w_2^2 a)
  \right]/\epsilon w_4$ and since $a \sim \mu^2$, we obtain $f_1(t) \approx
0$ below this time scale. For larger times $t \gg t_1$, the first term in the
parentheses can be ignored and for $\epsilon \to 0$, we obtain $f_1(t)
\sim e^{r_c t}$. Thus $z_2(t)=f_0(t) (1+f_1(t))$ increases as 
 \be
 z_2(t) 
\sim \begin{cases} 
(1+\epsilon w_4)^t~&,~t  \ll t_1 \\
(1+\epsilon w_4)^{\left[1+r_c/(w_4 \epsilon)\right] t}
     ~&,~t  \gg t_1 
\end{cases}
 \ee
Thus at times close to the end of phase I, $z_2$ increases at a
faster rate.  

To find the time $\tau_1$ at which phase I ends, we first calculate
$z_4(t)$ using $z_2(t)=f_0(t) (1+f_1(t))$ in (\ref{z2P1pos}). This yields
\be
z_4(t) \approx \frac{f_0(t+1) \left[1+f_1(t+1) \right]-\mu-w_2 f_0(t)
  \left[1+f_1(t)\right]+r w_2^2 f_0^2(t) \left[1+ 2 f_1(t) \right]}{r w_4}
\ee
On expanding $f_0(t+1)$ and $f_1(t+1)$ for
small $\epsilon$, we obtain 
\bea
f_0(t+1) &\approx&  (1+\epsilon w_4) f_0(t)\\
f_1(t+1) &\approx& f_1(t)+ \frac{1}{2} c \epsilon \alpha w_4 
(1+\epsilon w_4)^t \left[b+c (1+\epsilon w_4)^t \right]^{\alpha-1}
\eea
Substituting this in the above expression for $z_4(t)$ and using
$z_4(\tau_1)=1$, we find that $\tau_1$ is determined from the
following equation:
\be
\frac{(1-w_2) a}{2} y_1 (1+y_2)+ 
\frac{a r_c (c+2 a w_2^2)}{2} y_1^2 y_2=r w_4+\mu
\l{tau1comp}
\ee
where $y_1=(1+\epsilon w_4)^{\tau_1}$ and $y_2=(b+c
(1+\epsilon w_4)^{\tau_1})^\alpha$. 
As this equation is difficult to analyse, we consider only the fastest
growing term to obtain the following approximate equation for $\tau_1$:
\be
y_1^2 y_2 a r_c \left[c+2 a w_2^2\right] \approx 2 r w_4
\l{Atau1comp}
\ee

%--------------------------------------------------------------------
\noindent{\underline{\it Phase III.}} For $t > \tau_1$, the
populations $z_k$'s obey the equations (\ref{z4P3})-(\ref{z1P3}) as
for $e < 0$ and the corresponding solutions are given by
(\ref{z4P3neg})-(\ref{z1P3neg}) with $\tau_2$ replaced by $\tau_1$.   
For $w_2 > 1$, $z_2$ and $z_4$ grow exponentially
fast with their respective fitnesses but $z_1$ decays with time. The
rate of decline is determined by the ratio $w_2^2/(w_4-r)$. If
this ratio is larger than unity, $z_1 \sim (w_2^2/w_4)^t$ and as
$((w_4-r)/w_4)^t$ 
otherwise. For $w_2 < 1$, $z_2$ decays with time while $z_4$ continues
to grow. The remarks above for $z_1$ behavior when $w_2 > 1$ 
hold for $w_2 < 1$ case also.

%==========================================================================
%FIXATION TIME
%==========================================================================
\section{Fixation time}
\label{Time}

As seen in the last section, the unnormalised populations $z_k$'s vary
exponentially (or 
faster) with time so that the normalised population $x_4$ will reach unity 
asymptotically. Therefore, we define the fixation time $T$ as the time
when the population fraction $x_4(T)=1-\delta$ where $\delta \to 0$.
In terms of $z_k$'s, this condition gives
\be
z_1(T)+2 z_2(T)-\frac{\delta}{1-\delta} z_4(T) =0
\l{Tdef}
\ee
where $z_k(T)$ in the above equation is the population fraction in the
Phase III at $t=T$. Another reason why $\delta > 0$ is that for
$\delta=0$, the above equation cannot be 
satisfied as both $z_1$ and $z_2$ are always positive. 

For $r=0$ and $w_2 > 1$, since
$z_1 \approx 
1$, we can write $\delta z_4(T) \approx 2 z_2 (T)$ which gives 
\be
T \approx \frac{1}{\ln (w_4/w_2)} ~\ln \left[ \frac{2 (1-\delta) (w_4-w_2)
    (w_4-1)}{\mu \delta (w_4+w_2) (w_2-1)}\right]
\ee
which decreases monotonically as $w_4$ increases.

%==========================================================================
\subsection{No epistasis}

Since $z_4(t)=z_2^2(t)$ due to (\ref{z2e0}) and (\ref{z4e0}), the
condition (\ref{Tdef}) simplifies to give $z_2 (T)= (1+z_2 (T))
\sqrt{1-\delta}$ which leads to 
\be
T \approx \frac{1}{\ln w_2} \ln \left[1+\frac{(w_2-1)
    \sqrt{1-\delta}}{\mu (1-\sqrt{1-\delta})}  \right]
\l{Tnon}
\ee
For $w_2=2, w_4=4, \mu=10^{-6}$ and $\delta=0.01$, the above
 expression yields $T=27.56$ in excellent agreement with the result of
 our exact
 numerical iteration which gives the fixation time equal to $28$ for various
 values of $r$. 

%==========================================================================
\subsection{Negative epistasis}

Using (\ref{z4P3neg})-(\ref{z1P3neg}) at $t=T$ in (\ref{Tdef}), we have
\be
\frac{r w_2^2}{w_4 (r-e)} \left(\frac{w_2}{w_4}
\right)^{T-\tau_2}+2 -\frac{\delta}{1-\delta} \left(\frac{w_4}{w_2} \right)^{T-\tau_2}=0
\ee
Since the first term on the LHS is exponentially decaying, we 
neglect it to obtain
\be
T=\tau_2+ \frac{1}{\ln (w_4/w_2)} \ln \left[\frac{2
  (1-\delta)}{\delta} \right]
\l{Tneg}
\ee
where $\tau_2$ is given by (\ref{tau2}). As discussed in
Sec.~\ref{neg}, since $\tau_2$ decreases with
$r$, the fixation time $T$ is a decreasing function of recombination
probability $r$ when epistasis is negative. A comparison of the
analytical estimate with the exact numerical result for two sets of
parameters shows a good agreement (see Fig.~\ref{negtime}). 

\begin{figure}
\includegraphics[width=0.6 \linewidth,angle=270]{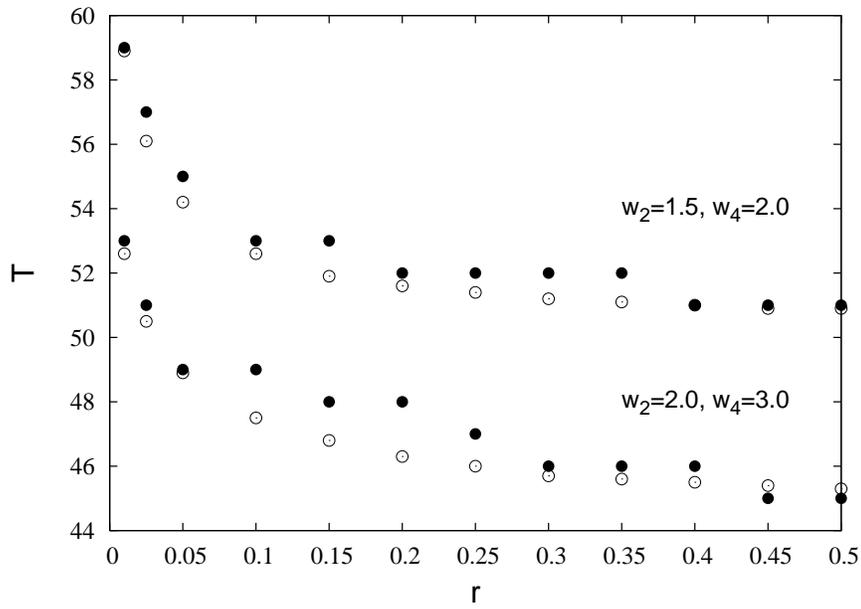}
\caption{Negative epistasis: Fixation time as a function of $r$
  obtained using exact 
  iteration ($\bullet$) and analytical results ($\circ$) given by
  (\ref{Tneg}) for two sets of fitnesses and $\mu=10^{-6},
  \delta=0.01$. The data for $w_2=2.0, w_4=3.0$ have been shifted by a
  constant by adding $12$.}
\l{negtime}
\end{figure}

%==========================================================================
\subsection{Positive epistasis}

Dividing both sides of (\ref{Tdef}) by $w_2$ and using
(\ref{z4P3neg})-(\ref{z1P3neg}), we find that the contribution due to
$z_1$ term can be neglected as it is exponentially decaying. This
gives 
\be
2 z_2(\tau_1) \approx \frac{\delta}{1-\delta} \left(\frac{w_4}{w_2} \right)^{T-\tau_1}
\ee
and hence 
\be
T = \tau_1+\frac{\ln \left[2 (1-\delta)/\delta \right]}{\ln (w_4/w_2)}+\frac{\ln z_2(\tau_1)}{\ln (w_4/w_2)}
\l{Tpos}
\ee
We first consider the $w_2 > 1$ case followed by $w_2 < 1$. 

%^^^^^^^^^^^^^^^^^^^^^^^^^^^^^^^^^^^^^^^^^^^^^^^^^^^^^^^^^^^^^^^^^^^^^^^
%w2 > 1
%^^^^^^^^^^^^^^^^^^^^^^^^^^^^^^^^^^^^^^^^^^^^^^^^^^^^^^^^^^^^^^^^^^^^^^^

\noindent \underline{$w_2 > 1$:} In the above expression,
$z_2(\tau_1)$ is given by (\ref{f0t}) and $\tau_1$ by 
(\ref{tau1lsse}) for $r < e/w_4$ and  (\ref{tau1rgtre}) for $r >
e/w_4$. Computing $T$ using these formulae in (\ref{Tpos}), we
obtain the fixation time as 
a function of $r$ shown in Fig.~\ref{postime} (open
circles) for fitness choices $w_2=1.25, w_4=1.8125, e/w_4 \approx
0.14$ and $w_2=1.25, w_4=2.5, e/w_4 =0.375$. The analytical data 
are seen to be in 
good agreement with the exact numerical results except in the vicinity
of $r=e/w_4$. Figure \ref{postime} also shows $\tau_1$ which displays a similar
behavior as $T$. 
We have already seen that the time $\tau_1$ increases linearly with
$r$ for $r \ll e/w_4$ but weakly for $r \gg e/w_4$. 
Thus the fixation time $T$ for $e > 0, w_2 > 1$ increases fast for
small $r$ but is weakly dependent on $r$ for $r > e/w_4$. 

%^^^^^^^^^^^^^^^^^^^^^^^^^^^^^^^^^^^^^^^^^^^^^^^^^^^^^^^^^^^^^^^^^^^^^^^
%w2 < 1
%^^^^^^^^^^^^^^^^^^^^^^^^^^^^^^^^^^^^^^^^^^^^^^^^^^^^^^^^^^^^^^^^^^^^^^^

\begin{figure}
\includegraphics[width=0.6 \linewidth,angle=270]{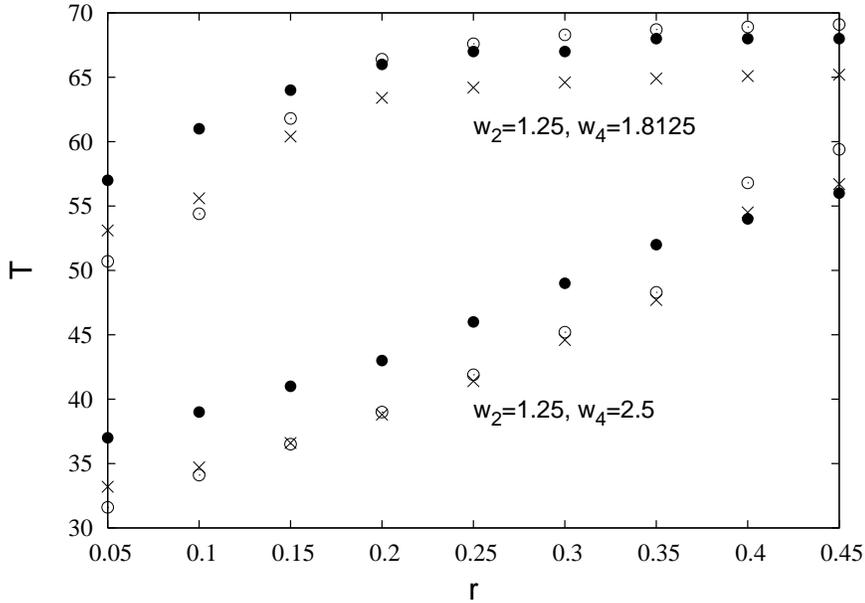}
\caption{Positive epistasis: Fixation time as a function of $r$
  obtained using exact 
  iteration ($\bullet$) and analytical result ($\circ$ and
  $\times$) given by 
  (\ref{Tpos}) and $\tau_1$ (upto a constant) respectively for two
  sets of fitnesses and $\mu=10^{-6}, \delta=0.01$.}
\l{postime}
\end{figure}

\begin{figure}
\includegraphics[width=0.6 \linewidth,angle=270]{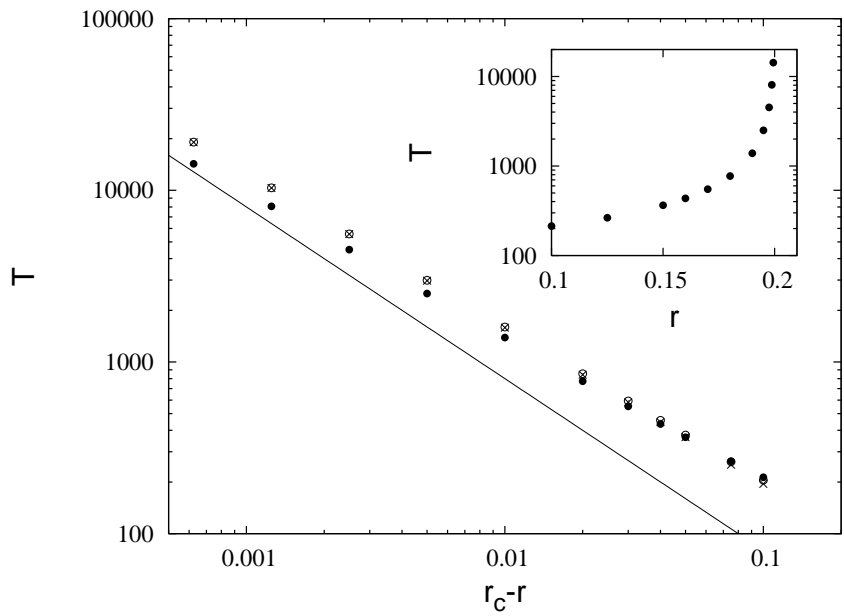}
\caption{Compensatory mutation: Fixation time as a function of $r$
  obtained using exact 
  iteration ($\bullet$) and analytical results ($\circ$ and
  $\times$) given by 
  (\ref{Tpos}) and (\ref{Atau1comp}) respectively for parameters
  $w_2=0.75, w_4=1.25, \mu=10^{-6}, \delta=0.01$ with $r_c=(w_4-1)/w_4$. The
  solid line has a slope equal to $-1$.}
\l{comptime}
\end{figure}

\noindent \underline{$w_2 < 1$:} The
inset of Fig.~\ref{comptime} shows that the 
fixation time diverges as $r$ approaches critical recombination
probability $r_c=(w_4-1)/w_4$. From (\ref{Tpos}), the fixation time
$T$ for $w_2 < 1$ can be calculated using $\tau_1$ from (\ref{tau1comp})
and $z_2(\tau_1)=f_0(\tau_1) (1+f_1(\tau_1))$. 
The time $T$ thus obtained (open circles) when plotted against $r_c-r$
is compared with the results from exact iteration in
Fig.~\ref{comptime} and shows a good agreement.  The approximate
$\tau_1$ obtained using (\ref{Atau1comp}) is also plotted which well
approximates the fixation time $T$. Therefore, it is sufficient to
analyse (\ref{Atau1comp}) in order to understand the behavior of the
fixation time as $r \to r_c$. For $\epsilon \to 0$, (\ref{Atau1comp}) 
can be written as 
\be
y_1^2 y_2 a r_c \left[c+2 a w_2^2\right] \approx 2 r_c w_4
\ee
On taking logarithms both sides, the above equation reduces to
\be
\frac{2 r_c^2 w_2^2 a}{(1-w_2) \epsilon} e^{{\tau_1} \epsilon w_4}
\approx \ln \frac{w_4 (1-w_2)}{(1-w_2+r_c w_4) w_2^2 a_2^2}
\ee
where we have neglected the linear term in $\tau_1$ as compared to the
exponential term in $\tau_1$. Writing $a={\tilde a}/\epsilon$, we
finally obtain 
\be
\tau_1 \approx \frac{1}{\epsilon w_4} \left[ \ln \ln \left(\frac{w_4 (1-w_2)
    \epsilon^2}{(1-w_2+r_c w_4) w_2^2 {\tilde a}^{2}} \right)-\ln
  \left(\frac{2 r_c^2 w_2^2{\tilde a}}{(1-w_2) \epsilon^2}\right) \right]
\l{tau1comp2}
\ee
which decays slower than $1/\epsilon$ (see Fig.~\ref{comptime}) due to
the logarithmic corrections.



%==========================================================================
%ASYMM
%==========================================================================
\section{Initial Condition with nonzero linkage disequilibrium}
\l{asymm}

So far, we have discussed the population dynamics starting with an
initial condition in which only one genotype has a nonzero
population. 
In this section, we consider the situation when a small finite
frequency at the intermediate loci is also present at $t=0$
i.e. $x_1(0) \neq 0, x_2(0)=x_3(0)=(1-x_1(0))/2$. As the
analytical method 
presented in the last sections assumes that all but one frequencies is
rare at a given time, it seems difficult to obtain analytical
results. Therefore we present numerical results to
show how the change in initial condition affects the fixation time. 

As shown in  Fig.~\ref{atime}, due to a nonzero population at
intermediate loci, the 
fixation time at a given $r$ is reduced as compared to the situation
when only the genotype $ab$ is present initially. For negative
epistasis (Fig.~\ref{atime}a), the trend in
the generalised situation appears similar to that discussed in
Fig.~\ref{negtime} in that the $T$ decreases slowly for small $r$ but
fast for large $r$. However for positive epistasis with $w_2 > 1$
(Fig.~\ref{atime}b), the fixation time remains roughly constant and
unlike Fig.~\ref{postime} does not increase. For compensatory
mutations, the fixation time shown in Fig.~\ref{actime} increases as
$r$ approaches a critical recombination rate and diverges slower than
$(r_c-r)^{-1}$. 

\begin{figure}
\includegraphics[width=0.39\linewidth,angle=270]{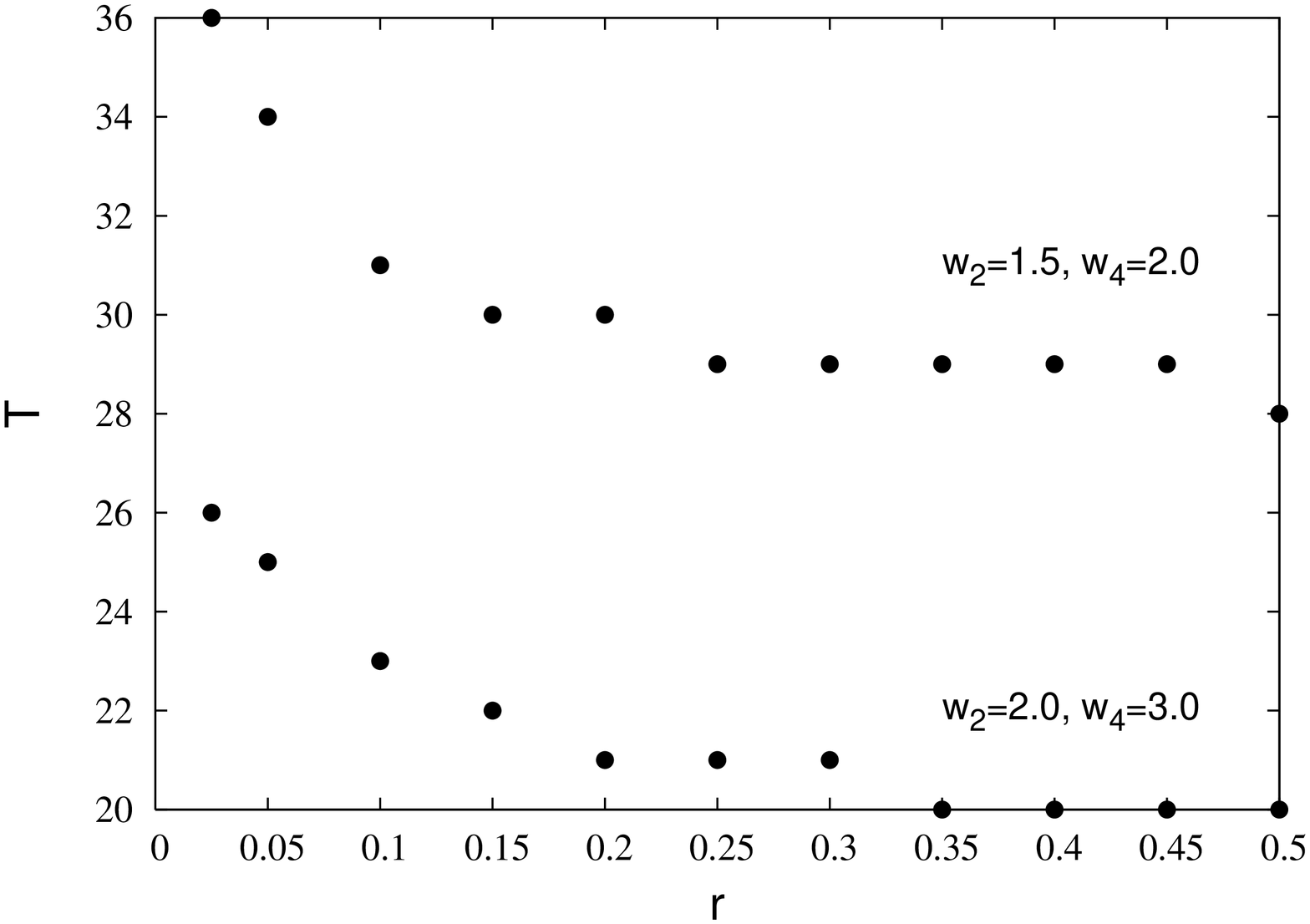}
\includegraphics[width=0.39 \linewidth,angle=270]{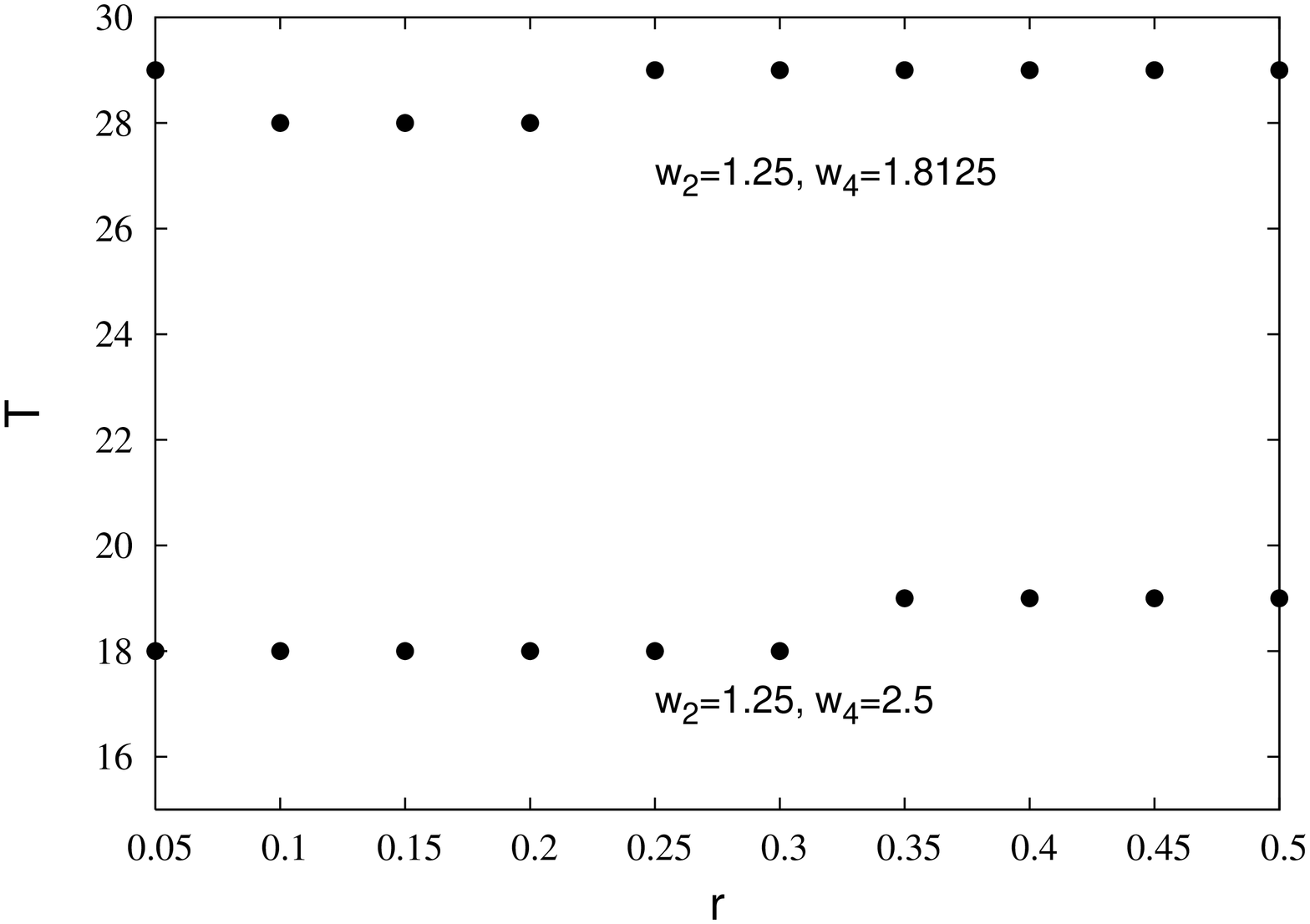}
\caption{Fixation time as a function of $r$
  obtained using exact iteration ($\bullet$) when $x_1(0)=0.95,
  x_2(0)=x_3(0)=0.025$ for the same parameters as in
  Figs.~\ref{negtime} and \ref{postime}.}
\l{atime}
\end{figure}

%==========================================================================
%CONCLUSIONS
%==========================================================================
\section{Conclusions}
\l{concl}

In this article, we have studied the dynamics of a 2 locus model in which the
population evolves deterministically under mutation, selection and
recombination. As the recombination process makes the equations
nonlinear, in general it is difficult to study such problems 
analytically. Here we have developed an analytical method to find the
fixation time to the best locus for various fitness schemes. 

\begin{figure}
\includegraphics[width=0.6\linewidth,angle=270]{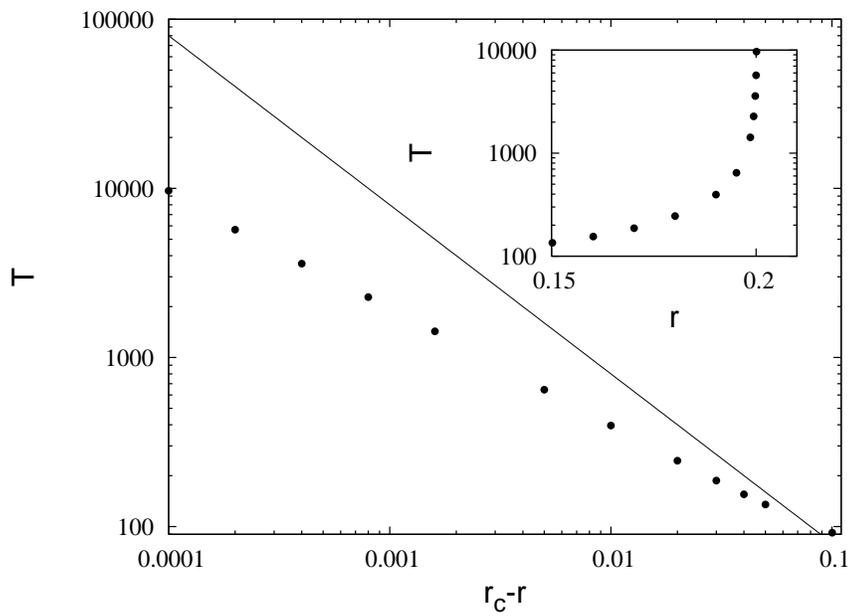}
\caption{Fixation time as a function of $r$
  obtained using exact iteration ($\bullet$) when $x_1(0)=0.95,
  x_2(0)=x_3(0)=0.025$ for the same parameters as in
  Figs.~\ref{comptime}. The numerically determined critical
  recombination rate $r_c \approx   0.2002$ and the solid line has a
  slope equal to $-1$. }
\l{actime}
\end{figure}

The fixation time $T$ is one of the measures for judging whether
recombination is beneficial for a population \cite{Feldman:1996}. If
the fixation time decreases with increasing recombination rate $r$, one
may deduce that recombining population has an advantage over an
asexual one. Our calculations show that when the epistasis parameter $e$ is
negative, the fixation time decreases fast for large $r$ which
suggests that high recombination rate may be beneficial for
populations with negatively epistatic fitness. 

In a fitness landscape with positive epistasis and fitness
increasing monotonically with mutational distance from the initial
sequence, the 
fixation time is shown to increase with recombination. As already
discussed in the Introduction, the result that $T$
increases with $r$ for positive epistasis is in qualitative agreement
with the expectation from the results of Eshel and Feldman
\cite{Eshel:1970}. Our analytical calculations show 
that the functional behavior of time $T$ depends on the ratio
$r w_4/e$. The fixation time is shown to 
increase linearly when $r \ll e/w_4$ but remains roughly constant for $r
\gg e/w_4$.

A compensatory mutation is said to occur when the fitness loss caused
by one mutation is remedied by its epistatic interaction with a second
mutation at a different site in the genome. For such a fitness scheme
in which the initial and final fitness hills are separated by a
fitness valley, it is known that an infinitely large population cannot
cross the intermediate valley beyond a critical recombination rate $r_c$ 
\cite{Crow:1965,Eshel:1970}. This implies that the fixation time
diverges as $r$ 
approaches $r_c$. Our exact numerical results for fixation time
when plotted against $r_c-r$ on a 
double logarithmic scale indicate a power law decay. Assuming that
the divergence is purely algebraic, a fit to the numerical data then
gives $T \sim (r_c-r)^{-0.83}$. However our calculation that  takes the 
nonlinearities into account shows that the divergence is actually 
a power law with logarithmic corrections. 

Here we have focused on the evolution of deterministic population but it is
important to include drift effects as the real populations have a
finite size $N$. 
The finite population problem with compensatory mutation has been studied
in certain parameter regimes using simulations and within a
diffusion approximation.  
The analytical calculations of
\cite{Stephan:1996,Iizuka:1996,Higgs:1998} and numerical 
simulations of \cite{Michalakis:1996, Phillips:1996} for a two-locus model with
compensatory mutation indicates that for fixed $s$ and $N$, a finite
population manages to reach the best locus for any $r$ and the
fixation time increases with $r$ as seen for infinite population. However 
it is not clear how the population approaches the infinite $N$ 
limit. For  $e > 0, w_2 > 1$, it is found numerically there exits a
critical epistasis value below which the fixation time
decreases \cite{Otto:1994}. An analytical  
understanding of such interesting aspects of the interplay between 
recombination and drift remains an open problem.

\begin{table}
\begin{center}
\begin{tabular*}{0.9\textwidth}{@{\extracolsep{\fill}}lll}
\hline
Title & Conditions& Fixation time $T$ \\
\hline
$e=0$ & $r > 0$ & Independent of $r$ (\ref{Tnon})\\
$e < 0, w_2 > 1$ & $r > e$ &  Decreases with $r$ (\ref{Tneg}) \\ 
$e > 0, w_2 > 1$ & $r < e/w_4$ & Increases linearly with $r$ (\ref{tau1rle})\\
   & $r > e/w_4$ &  Increases weakly with $r$ (\ref{tau1rgtre})\\
$e > 0, w_2 < 1$& $r < r_c$ & Diverges with $r_c-r$ (\ref{tau1comp2})\\
 & $r > r_c$ & Infinite \\
\hline
\end{tabular*}
\caption{Table summarising the dependence of fixation time $T$ on
  recombination probability $r$ for various choices of epistasis $e$. }
\label{summary}
\end{center}
\end{table}

Acknowledgement: The author is grateful to J. Krug for introducing her
to this problem and useful discussions. She also thanks S.-C. Park for
comments on the manuscript and 
KITP, Santa Barbara for hospitality where a part of this work was done. 

%==========================================================================
%==========================================================================

\end{document}